# TWO-DIMENSIONAL ANHARMONIC CRYSTAL LATTICES: SOLITONS, SOLECTRONS, AND ELECTRIC CONDUCTION

**Manuel G. Velarde, Werner Ebeling, Alexander P. Chetverikov,** *Instituto Pluridisciplinar, Universidad Complutense de Madrid (Spain)*

**Abstract**
Reported here are salient features of soliton-mediated electron transport in anharmonic crystal lattices. After recalling how an electron-soliton bound state (solectron) can be formed we comment on consequences like electron surfing on a sound wave and *ballistic* transport, possible percolation in 2d lattices, and a novel form of electron pairing with strongly correlated electrons both in real space and momentum space.

## 1. Introduction

Electrons, holes, or their dressed forms as "quasiparticles", in the approach introduced by Landau (Landau 1933, Kaganov 1979), play a key role in transferring charge, energy, information or signals in technological and biological systems (Slinker 2011). Engineers have invented ingenious methods for, e.g., long range electron transfer (ET) such that an electron and its "carrier", forming a quasiparticle, go together all along the path hence with space and time synchrony. Fig. 1 illustrates the simplest geometry between a donor (D) and an acceptor (A). Velocities reported are in the range of sound velocity which in bio-systems or in *GaAs* layers are about Angstrom/picosecond (Km/s). Such values are indeed much lower than the velocity of light propagation in the medium. Thus at first sight, leaving aside a deeper discussion concerning specific purposes (Pomeau 2007, Chetverikov 2012a), controlling electrons seems to be more feasible with sound (or even supersonic) waves than with photons. Electron *surfing* on an appropriate highly monochromatic, quite strong albeit linear/harmonic wave has recently being observed (Hermelin 2011, McNeil 2011). Earlier the present authors have proposed the *solectron* concept as a new "quasiparticle" (Velarde 2005, 2006, 2008a, 2008b, 2010a, 2010b, Chetverikov 2006a, 2009, 2010, 2011a, 2011b, Cantu Ros 2011, Hennig 2006, Ebeling 2009) encompassing lattice anharmonicity (hence invoking nonlinear elasticity beyond Hooke's law) and (Holstein-Fröhlich) electron-lattice interactions thus generalizing the polaron concept and quasiparticle introduced by Landau and Pekar (Landau 1933, Pekar 1954, Devreese 1972). Anharmonic, generally supersonic waves are naturally robust due to, e.g., a balance between nonlinearity and dispersion (or dissipation). In the following Sections we succinctly describe some of our findings and predictions for one-dimensional (1d) crystal lattices for which exact analytical and numerical results exist (Section 2) and, subsequently, for two-dimensional (2d) lattices for which only numerical results are available (Sections 3, 4 and 5). Comments about theory and experiments are provided in Section 6 of this text.

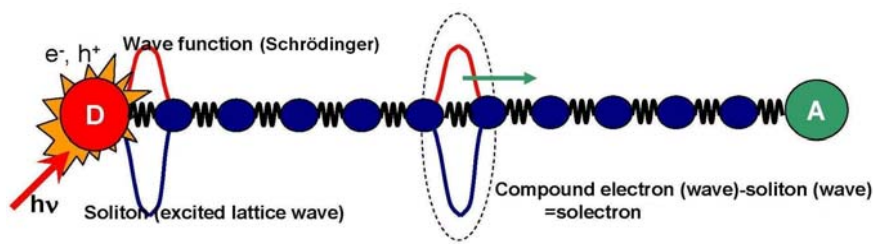

*Figure 1: Electron transfer from a donor (D) to an acceptor (A) along a 1d crystal lattice. The springs mimic either harmonic interactions or otherwise. In this text they are assumed to correspond to (anharmonic) Morse potentials. The figure also illustrates the electron-soliton bound state (solectron) formation. Depending on the material ways other than photoexcitation at the donor site could lead to the same consequences.*

## 2. Soliton assisted electron transfer in 1d lattices.

Although the basic phenomenological theory exists (Marcus 1985) yet long range ET (beyond 20Å) in biomolecules is an outstanding problem (Gray 2003, 2005). Recent experiments by Barton and collaborators (Slinker 2011) with synthetic DNA show an apparent *ballistic* transport over 34 nm for which no theory exists. Let us see how we can address this question building upon our *solectron* concept (Fig. 1).

We consider the 1d-crystal lattice with *anharmonic* forces described by the following Hamiltonian

$$H_{lattice} = \sum_n \left\{ \frac{Mv_n^2}{2} + D(1 - \exp[-B(x_n - x_{n-1} - \sigma)])^2 \right\}, \tag{1}$$

where $x_n$, $v_n$, $M$, $D$, $B$ and $\sigma$ denote, respectively, space lattice coordinates/sites, lattice particle/unit velocities, unit masses (all taken equal), the potential depth or dissociation energy of the Morse potential (akin to the 12-6 Lennard-Jones potential), lattice stiffness constant and interparticle equilibrium distance or initial lattice spacing. For our purpose here we introduce suitably rescaled relative lattice displacements, $q_n = B(x_n - n\sigma)$. Around the minimum of the potential well we can define $\omega_0 = (2DB^2/M)^{1/2}$ as the linear (harmonic) vibration frequency. For biomolecules like azurin $\omega_0 = 10^{13} s^{-1}$, $M \approx 100$ amu. Then for $D = 0.1\,eV$ ($D = 1.0\,eV$) we can set $B = 2.3\,Å^{-1}$ ($B = 0.72\,Å^{-1}$) (Velarde 2010b).

If an *excess* electron is added to the lattice we can take it in the tight binding approximation (TBA) and hence

$$H_{electron} = \sum_n E_n C_n^* C_n - \sum_n V_{n,n-1}(q_k)\left(C_n^* C_{n-1} + C_n C_{n-1}^*\right), \tag{2}$$

with $n$ denoting the lattice site where the electron is placed (in probability density, $w_n = |C_n|^2$, $\sum_n w_n = 1$). We want to emphasize the significance of hopping in the transport process relative to effects due to onsite energy shifts and hence we assume $E_n = E_0$ for all $n$ save those referring to D and A. The quantities $V_{n,n-1}$ belong to the transfer matrix or overlapping integrals. They depend on actual relative lattice displacements, and we can set (Slater 1974)

$$V_{n,n-1} = V_0 \exp[-\alpha(q_n - q_{n-1})], \tag{3}$$

where $V_0$ and $\alpha$ account for the electron-lattice coupling strength. Accordingly, $\tau = V_0/\hbar\omega_0$ provides the ratio of the two dynamical time scales (electronic over mechanical/sound).

From (1)-(3) follow the equations of motion in suitable dimensionless form:

$$\frac{d^2 q_n}{dt^2} = \left[1 - e^{(q_n - q_{n+1})}\right] e^{(q_n - q_{n+1})} - \left[1 - e^{(q_{n-1} - q_n)}\right] e^{(q_{n-1} - q_n)} +$$
$$+ 2\alpha V \left[\text{Re}\left(C_{n+1} C_n^*\right) e^{\alpha(q_n - q_{n+1})} - \text{Re}\left(C_n C_{n-1}^*\right) e^{\alpha(q_{n-1} - q_n)}\right] \tag{4}$$

$$\frac{dC_n}{dt} = i\tau \left[C_{n+1} e^{\alpha(q_n - q_{n+1})} + C_{n-1} e^{\alpha(q_{n-1} - q_n)}\right]. \tag{5}$$

It is worth recalling that if rather than the Morse potential (1) we use a similar potential introduced by Toda the lattice dynamic problem defined by Eqs. (4) in the absence of the added electron ($\alpha = 0$) is exactly solvable (Toda 1989, Chetverikov 2006b). Thus we know analytical expressions for lattice motions and, moreover, for the thermodynamics/statistical mechanics (including specific heats, dynamic structure factor, etc.) of such 1d many-body problem. For the Morse potential (1) it has been numerically shown that no significant differences exist for lattice motions and other physical quantities (Dancz 1977, Rolfe 1979). Temperature can be incorporated in the dynamics by adding to Eqs. (4) Langevin sources by using an appropriate heat bath (delta-correlated Gaussian white noise) and using Einstein's relation between noise strength and temperature. To avoid redundancy we illustrate this point in Sect. 3.

The implementation of the scheme shown in Fig. 1 is one prediction with velocities in the sonic and supersonic range. Fig. 2 illustrates the possibility using Eqs. (4)-(5) of extracting an electron placed in a potential well in the 1d Morse lattice by a generally supersonic soliton. For the geometry of Fig. 1 we can use it to estimate the *ballistic* process time lapse to go from the donor to the acceptor. For the computation with a lattice of $N = 100$ units the well is assumed Gaussian of depth $|E|$ (in units of $\hbar\omega_0$) with $E < O$ localized at site 50. The soliton initially spans a few lattice sites (two or three) excited at site 40. If the well depth is shallow enough the extraction is ensured up to 100% whereas if the well is too deep no extraction occurs. Needless to say extraction is possible with probability varying from zero to unity as the well depth is decreased. Time lapse from *D* to *A* is obtained by simply dividing length over soliton speed. Illustration is provided in Fig. 3 where "$\ell$" (see Fig. 1) accounts for the distance travelled (in principle from *D* to an appropriately placed acceptor *A*). Comparison is provided between the *ballistic* case and other possibilities like *diffusion*-like transport with thermally (hence randomly) excited solitons (Chetverikov 2010) and *tunneling* transport (Chetverikov 2012b).

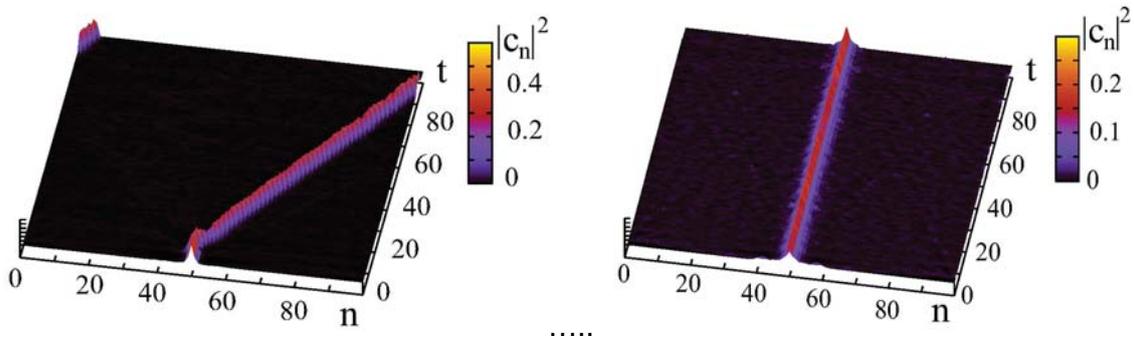

…..

*Figure 2: Extraction of an electron from a potential well (a donor) and ballistic transport to an acceptor observed using the electron probability density $|C_n|^2$. Left panel: shallow well $|E| = 10$, extraction 100%. Right panel: deep well $|E| = 18$, no extraction. Parameter values: α = 1.75, $V_0$ = 0.35 and τ = 10.*

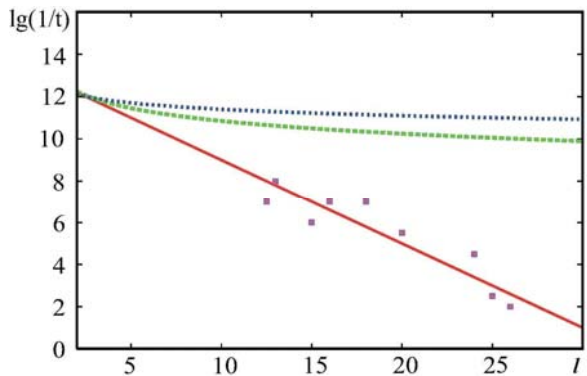

*Figure 3: Logarithm of reciprocal time lapse (in seconds) which an electron bound to a soliton needs to travel a distance l (in Angstrom) for the geometry of Fig. 1. The upper dotted (blue) curve corresponds to a sound velocity of 17 Angstrom/ps, illustrating a ballistic transport. The second dotted (green) curve from above shows the reciprocal time needed if the electron hops stochastically between thermally excited solitons. The bottom solid line embraces data illustrating a tunneling process. The dots are reciprocal times measured for natural bio-molecules (Gray 2003, 2005). The transfer times found for synthetic DNA are much shorter (Slinker 2011) bearing similarity to our model findings –upper dotted (blue) line– for solectron transfer.*

## 3. Two-dimensional crystal lattices

Recently, two groups of experimentalists have observed how an electron can "surf" on a suitably strong albeit linear, highly monochromatic sound wave (in *GaAs* layers at 300 mK). Sound demands lattice compressions and hence is accompanied by electric/polarization fields which for piezoelectric crystals integrate to macroscopic level. Our theoretical *solectron* approach targets sound-wave electron surfing due to nonlinear soliton excitations in 2d-anharmonic crystal lattice layers, with velocities ranging from supersonic to sub-sonic (Chetverikov 2006b, Hennig 2006, Makarov 2006). We do not pretend here to explain the *GaAs* experimental results. We simply wish to point out that appropriate sound waves in suitable nonlinear crystalline materials, could provide long range ET in 2d, with sonic or supersonic velocities for temperatures much higher than that so far achieved in experiments.

In 2d the Morse potential needs to be truncated to avoid overcounting lattice sites. Then using complex coordinates $Z = x + iy$, where $x$ and $y$ are Cartesian coordinates, the equations of motion replacing Eqs. (4) are

$$\frac{d^2 Z_n}{dt^2} = \sum_k F_{nk}(|Z_{nk}|)z_{nk} + \left[-\gamma \frac{dZ_n}{dt} + \sqrt{2D_v}(\xi_{nx} + i\xi_{ny})\right], \qquad (6)$$

with $F_{nk}(|Z_{nk}|) = -(dV/dr)_{r=|Z_{nk}|}$, $z_{nk} = (Z_n - Z_k)/|Z_n - Z_k|$. In Eqs. (6) we have incorporated thermal effects. The quantities $\gamma$ (friction coefficient), $D_v$ (diffusion coefficient) and the $\xi_s$ (noise generators) characterize the Gaussian noise. $D_v = k_B T \gamma / M$ is Einstein's relation with $k_B$, Boltzmann constant.

To illustrate lattice motions we consider each lattice unit as a sphere representing the core electron Gaussian distribution at the corresponding site: $\rho(Z,t) = \sum_{|Z-Z_i(t)|<1.5} \exp\left(-|Z - Z_j(t)|^2 / 2\lambda^2\right)$ with $\lambda$ a parameter. Thus overlapping of two such Gaussians permit to "detect" the expected "mechanical" compression of two lattice units as Fig. 4 illustrates (Chetverikov 2011c, 2011d). The evolution of the electron follows Eq. (5) for the 2d lattice geometry.

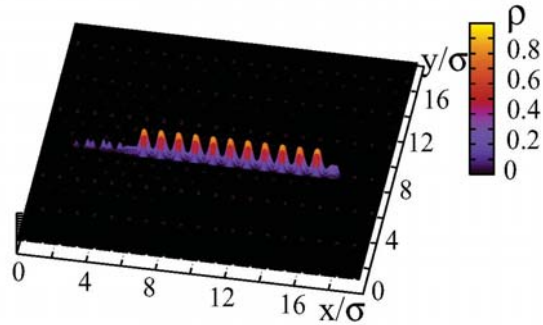

*Figure 4: Cumulative sequence of snapshots using $\rho(Z,t)$ to track a soliton running along the x-axis of a triangular lattice using Eqs. (4), (6). Parameter values: $B\sigma = 4$, $T = 0.001$.*

## 4. Two-dimensional crystal lattices. Pauli's master equation approach

Continuing with the 2d case, we now consider an alternative approach to using the Schrödinger equation (5). We shall consider how transport is achieved following Pauli's master equation approach (Ebeling 2009). Eq. (2) is now considered with $V_{nn'}(Z_{n,n'})$. The energy levels are taken in the *polarization* approximation $E_n = E_0 - \sum_n \left\{ U_e h^4 \Big/ \left[ |Z_{n,n'}|^2 + h^2 \right]^2 \right\}$, where $U_e$ is the electric potential strength and $h$ defines the range of the electric field polarization interaction. Rather than relying on the Schrödinger description of the TBA we follow Pauli's master equation approach with transition probabilities

$$W_{n,n'} = (t_0^2/\hbar)\exp[-2\alpha|Z_{n,n'}|]E(n,n';\beta), \qquad (7)$$

$$\frac{dw_n}{dt} = \sum [W_{nn'}w_{n'} - W_{n'n}w_n], \qquad (8)$$

where $E(n,n';\beta) = 1$ if $E_n < E_{n'}$ and $E(n,n';\beta) = \exp[-\beta(E_n - E_{n'})]$ if $E_n > E_{n'}$. Eqs. (7)-(8) are solved with Eqs. (6) to obtain the electron probability density $w_n(t)$ neglecting the feedback of the electron on the lattice dynamics.

Figs. 5 and 6 illustrate electron and solectron evolution along a 2d lattice. Fig. 5 refers to electron taken alone while Fig. 6 illustrates how, after switching-on the electron-lattice interaction, the soliton from Fig. 4 is able to trap the electron from Fig. 5 and after forming the solectron transports charge along the lattice (see also Velarde 2008a).

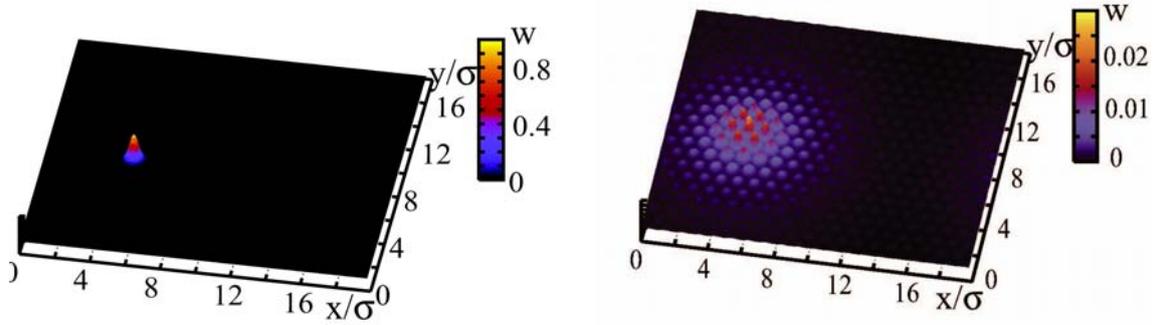

*Figure 5: An electron alone placed at a given lattice site (left panel). The quantity $w$ here accounts for the probability density (otherwise $|C_n|^2$ in Fig. 3). As time progresses the electron spreads over the slightly heated lattice $(T = 0.002D)$ following Pauli's equation from the initial condition (left panel) to a subsequent time instant (right panel).*

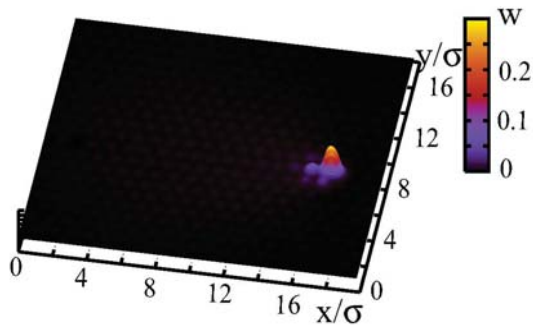

*Figure 6: Solectron formation and eventual evolution when the electron-phonon (here electron-soliton) interaction is switched-on ($\alpha \neq 0$ in the corresponding to Eqs. (4) for the 2d case). We see that the electron is trapped by a soliton (like that of Fig. 4) thus forming the solectron which transfers the electron probability density without spreading at variance with the result illustrated in Fig. 5 (right panel).*

## 5. Percolation and other features in 2d-lattices

Solitons can be excited in a crystal lattice by several actions. One is to add finite momentum to a group of nearby lattice units, another is by heating the crystal all-together. Then one expects quite many excitations including phonons and solitons randomly appearing along the 2d lattice and having finite life times thus leaving finite-length traces. Fig. 7 illustrates thermal excitations leading to spots of instant electron density $n_e(x,y;t)$ due to higher than equilibrium electric/polarization field maxima. Here in the simplest Boltzmann approximation

$$n_{el}(Z/x,y;t) = \{\exp-[U(Z,t)/k_B T]/n_{el}^0\}, \qquad (9)$$

With $n_{el}^0$ the normalizing factor, $n_{el}^0 = \int \exp[-U(Z,t)/k_B T]dZ$.

Only at temperatures high enough one expects a distribution of "local" spots permitting in kind of zig-zag the occurrence of an "infinite" path thus percolating from side to side of the 2d lattice (Chetverikov 2009, 2011a). Indeed by increasing temperature one increases the significance as well as the "density" of soliton excitations/traces. If percolation does occur by adding an excess electron and playing with an external field we have a novel way of one-sided electric conduction mediated by the solitons. We have just explored this possibility but have not yet been able to draw conclusions about the scaling laws of the process. On the other hand since percolation is expected as a second-order phase transition it seems worth investigating the possible connection with the pseudo-gap transition observed in such superconducting materials as cuprates.

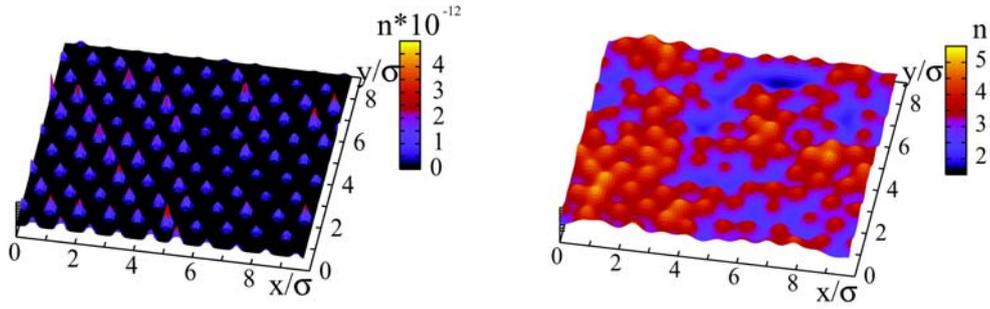

*Figure 7: Towards percolation. Instantaneous space distribution of electron probability density $n(x,y)$ associated to lattice solitons (sound) in a triangular Morse lattice (N = 100) at, respectively, low (T = 0.02 D) (left panel) and high (T = 0.4 D) (right panel) temperatures. The latter exhibits an almost percolating path. Parameter values: B$\sigma$ = 3, $U_e$ = 0.4D, h = 0.7$\sigma$.*

## 6. Concluding remarks

We have illustrated how lattice solitons arising from *finite* amplitude compression-expansion longitudinal motions bring sound and also create electric polarization fields (Landau 1933). The latter are able to trap charges and provide long-range ET in a wide range of temperatures (up to e.g. 300K for bio-molecules). Such "sound" waves could exhibit subsonic, sonic or supersonic velocity, whose actual value depends on the strength of the electron-phonon/soliton interaction. Noteworthy is that such interaction and subsequent electric transport, in the most general case, embraces a genuine *polaron* effect (Pekar 1954, Devreese 1972) and also a genuine soliton/solectron effect (Cantu Ros 2011). For piezoelectric materials like *GaAs* that sound waves can transport electrons there is now experimental evidence (Hermelin 2011, McNeil 2011). This was achieved by means of strong albeit linear/infinitesimal, highly monochromatic waves appropriately creating the electric/polarization field that due to the specificity of the crystal symmetry and other features integrate to macroscopic level. These experiments done at 300 mK due to quantum limitations imposed to the set-up provide hope for similar long-range ET at "high" temperatures. Indeed the limitations are only due to the electron entry and exit/detector gates. The *solectron* theory predicts such a possibility in appropriate non-linearly elastic crystal materials capable of sustaining lattice solitons. Recent experiments using synthetic DNA (Slinker 2011) show a kind of *ballistic* ET over 34 nm which as Fig. 3 illustrates bears similarity with a prediction of our *solectron* theory (Velarde 2008a).

In 2d crystal lattices the *solectron* theory predicts the possibility of percolation as a way of long range charge transport when the material is heated up to the range of robustness/stability of lattice solitons, as Fig. 7 illustrates. Work remains to be carried out to assess the corresponding percolation scaling laws.

Finally, we have recently shown that the *solectron* theory offers a new way of electron *pairing* by having two electrons strongly correlated (both in real space and in momentum space with due account of Pauli's exclusion principle and Coulomb repulsion using Hubbard's local approximation) due to their trapping by lattice solitons (Velarde 2006, 2008c, 2011a, 2011b, 2011c, Hennig 2008, Brizhik 2012). This feature shows the quite significant role played by the lattice dynamics well beyond the role played in the formation of Cooper pairs (in momentum space) underlying the BCS theory (Cooper 2011) or in the bipolaron theory (Alexandrov 2007) and much in the spirit of Fröhlich approach to the problem unfortunately using a harmonic lattice Hamiltonian at a time before (lattice) solitons were known (Fröhlich 1950, 1952, 1954a, 1954b, Zabusky 1965, 2005, Toda 1989; see also Nayanov 1986). Incidentally, Einstein (1922) was the first who used the concept of molecular conduction chains trying to understand superconduction. Thus it is reasonable to expect that a soliton-mediated Bose-Einstein condensation could take place in appropriate 2d *anharmonic* crystal lattices well above absolute zero. This is yet to be shown.

## Acknowledgments

The authors are grateful to A. S. Alexandrov, E. Brändäs, L. Brizhik, L. Cruzeiro, F. de Moura, J. Feder, D. Hennig, R. Lima, R. Miranda, R. McNeil, G. Röpke and G. Vinogradov for enlightening discussions. This research has been sponsored by the Spanish Ministerio de Ciencia e Innovación, under Grants EXPLORA FIS2009-06585 and MAT2011-26221.